\documentclass{emulateapj}

\shorttitle{Re-examination of the Expected Gamma-Ray Emission of
	Supernova Remnant SN~1987A}
\shortauthors{E.G. Berezhko, L.T. Ksenofontov \& H.J. V\"olk}

\newcommand{\gr}{$\gamma$-ray \,}
\newcommand{\grs}{$\gamma$-rays \,}

\begin{document}

\title {Re-examination of the Expected gamma-ray emission of supernova remnant 
SN~1987A}

\author{E.G.~Berezhko\altaffilmark{1},
        L.T.~Ksenofontov\altaffilmark{1},
and     H.J.~V\"olk\altaffilmark{2}
}
\altaffiltext{1}{Yu. G. Shafer Institute of Cosmophysical Research and Aeronomy,
                 31 Lenin Avenue, 677980 Yakutsk, Russia}
\altaffiltext{2}{Max Planck Institut f\"ur Kernphysik,
                Postfach 103980, D-69029 Heidelberg, Germany}

\email{ksenofon@ikfia.sbras.ru}

\begin{abstract}
  A nonlinear kinetic theory, combining cosmic-ray (CR) acceleration in
  supernova remnants (SNRs) with their gas dynamics, is
  used to re-examine the nonthermal properties of the remnant of SN~1987A
  for an extended evolutionary period of 5--50 yr. This spherically symmetric
  model is approximately applied to the different features of the SNR which
  consist of (i) a blue supergiant wind and bubble, and (ii) of the swept-up
  red supergiant (RSG) wind structures in the form of an H~II region, an
  equatorial ring (ER), and an hourglass region. The RSG wind involves a mass
  loss rate that decreases significantly with elevation above and below the
  equatorial plane. The model adapts recent three-dimensional hydrodynamical
  simulations by Potter et al. in 2014 that use a significantly smaller ionized
  mass of the ER than assumed in the earlier studies by the present
  authors. The SNR shock recently swept up the ER, which is the densest
  region in the immediate circumstellar environment. Therefore, the expected
  gamma-ray energy flux density at TeV energies in the current epoch has
  already reached its maximal value of $\sim 10^{-13}$~erg cm$^{-2}$ s$^{-1}$.
  This flux should decrease by a factor of about two over the next 10 years.
\end{abstract}

\keywords{acceleration of particles --- ISM: individual objects (SN 1987A) --- 
ISM: supernova remnants --- X-rays: individual (SN 1987A) --- gamma rays: ISM}


\section{Introduction}
The supernova (SN) that occurred in 1987 in the nearby Large Magellanic Cloud
was the first object of its kind whose evolution in the radio to X-ray
range has been resolved as a function of time. The study of SN~1987A continues
and includes, in particular, extensive observations in very high energy (VHE;
$E>100$~GeV) \grs \citep[e.g.][]{hess15}.

The present work is a critical re-examination and extension of the studies by
\citet{bk00,bk06} and \citet{bkv11}(referred to as BKV11 in the following)
concerning the properties of the nonthermal emission of SN 1987A. As in those
previous investigations, our framework is a nonlinear kinetic theory of
cosmic-ray (CR) acceleration in SN remnants (SNRs). This theory couples the particle 
acceleration process with the hydrodynamics
of the thermal gas \citep{byk96} and connects it with the gamma-ray emission
\citep[e.g.][]{bv00}. The application of this theory to individual SNRs
\citep[see][for reviews]{vlk04,ber05,ber08,ber14} has demonstrated its
ability to explain the observed SNR properties and radiation
spectra. Combining the theoretical model with the {\it observed} synchrotron
spectra predicts new effects like the large degree of magnetic field
amplification that leads to the observed concentration of the highest-energy
electrons in a very thin shell just behind the forward shock into the
circumstellar medium (CSM).

  The evolution of radio and X-ray emission at earlier times, also implied
  in the present paper, has been described by, e.g. BKV11, who showed, in
  particular, that the evidence for strong shock modification comes primarily
  from radio data.

  Efficient acceleration of the CR proton component is needed to produce
  significant shock modification leading to a soft and concave CR electron
  spectrum in SN 1987A, which well fits the observed nonthermal radio
  emission and X-ray spectra if the downstream magnetic field strength
  $B_\mathrm{d}$ is as high as $\approx 15$~mG. Such a high field leads to
  significant electron synchrotron losses that cut off the high-frequency X-ray
  part of the synchrotron spectrum, consistent with X-ray observations. The
  required magnetic field strength can presumably be attributed to its
  nonlinear amplification near the SNR shock by the CR acceleration process
  itself \citep[][]{bell04}.

As a consequence of the efficient production of the CR nuclear component, which
is accompanied by strong magnetic field amplification, SN~1987A is expected to
be a potential source of \grs for the H.E.S.S. instrument. However, the
flux of TeV emission predicted in BKV11 exceeds the upper
limit obtained by H.E.S.S. \citep{hess15}. It is suggested here that this
inconsistency is the result of an overestimate of the CSM gas density, in
particular, of the mass of the equatorial ring (ER).

The present work uses the same canonical values \citep[e.g.][]{mccray93} for
the stellar ejecta mass, $M_\mathrm{ej}$, distance, $d$, hydrodynamic explosion
energy, $E_\mathrm{sn}$, and ejecta velocity distribution as in BKV11, but takes
into account the detailed radio continuum observations by \citet{ng13} and
\citet{zanardo14}. In particular, the present work makes use of
recent extensive CSM modeling in terms of three-dimensional
  hydrodynamical simulations by \citet{potter14} (in the sequel referred to as
Potter14). An approximate prediction of the future nonthermal emission from
SN~1987A for the coming decades is given. The
densest part of the CSM, which lies within the range $|\theta|<20^{\circ}$ of
the elevation angle $\theta$ around the equatorial plane, is then considerably
less dense than what was adopted in BKV11. This is the main reason for the
re-examination of the CR production in SN~1987A and the associated nonthermal
emission.

\section{SN~1987A and its circumstellar environment}

To study the propagation of the SN shock through the CSM, the results of
Potter14 for the angular range $|\theta|<20^{\circ}$ relative to the equatorial
plane are used\footnote{The equatorial-to-polar density ratio is 20:1
  \citep{bl93}.}. The most efficient CR and nonthermal emission production
presumably takes place within this region. This is roughly consistent with
radio observations \citep{ng13}. The adopted radial profile of the gas number
density $N_\mathrm{g}=\rho/m_\mathrm{p}$ in this region is represented in
Figure~\ref{f1}. Within the selected elevation range, it consists of several
different morphological structures: (i) the wind bubble of the blue supergiant
(BSG) progenitor star \citep{chf87} at $r<R_\mathrm{C} = 4.5\times 10^{17}$~cm
with gas density $N_\mathrm{g}=0.29$~cm$^{-3}$, (ii) the H~II region
\citep{chd95} at $R_\mathrm{C}<r<R_\mathrm{HG}=8\times10^{17}$~cm with
$N_\mathrm{g}=280$~cm$^{-3}$, (iii) the so-called hourglass region at
$R_\mathrm{HG}<r<R_\mathrm{W}=1.5\times10^{18}$~cm with $N_\mathrm{g}=10$~cm$^{-3}$, and
(iv) the free red supergiant (RSG) wind region at $R>R_\mathrm{W}$ with
$N_\mathrm{g}=10(r/R_\mathrm{W})^3$~cm$^{-3}$; the properties of
structures (iii) and (iv) directly follow from Potter14. Within the smaller
elevation angle region of $|\theta|<4.5^{\circ}$, the same radial profile
includes the equatorial ring inside the H~II region (see Figure~\ref{f1}). Its
gas number density is chosen here to be distributed according to the relation
\begin{equation}
N_\mathrm{g}=N_\mathrm{gm} \exp[-(r-R_\mathrm{ER})^2/l_\mathrm{ER}^2],
\end{equation}
where $N_\mathrm{gm}\approx M_\mathrm{ER}/(4\pi^{3/2}m_p
R_\mathrm{ER}^2l_\mathrm{ER})$ is the central (maximal) density of the ER, and
$M_\mathrm{ER}$, $R_\mathrm{ER}$, and $l_\mathrm{ER}$ denote the total mass,
radius, and width of the ER, respectively. Below, the values
$M_\mathrm{ER}=0.058M_{\odot}$, $R_\mathrm{ER}=6.4\times10^{17}$~cm, and
$l_\mathrm{ER}=0.12R_\mathrm{ER}$ are used. These parameter values are taken in
order to fit the observed shock and radio-emission dynamics. They turn out 
to be consistent with the values used by Potter14.

The CSM described above differs significantly from the CSM adopted in BKV11.
First of all, the ER mass, albeit that of the ionized material only, 
$M_\mathrm{ER}=0.058M_{\odot}$ \citep{mlg10}, is considerably smaller
than that used by BKV11 ($M_\mathrm{ER}=0.5 M_{\odot}$). Second, the CSM behind the
ER is less dense. These factors, as demonstrated below, lead to a considerable
reduction of the expected nonthermal emission, in particular, of the \gr
emission, compared with the results of BKV11.

\section {Particle acceleration model}

\subsection{Shock Approximation}

\begin{figure}
\plotone{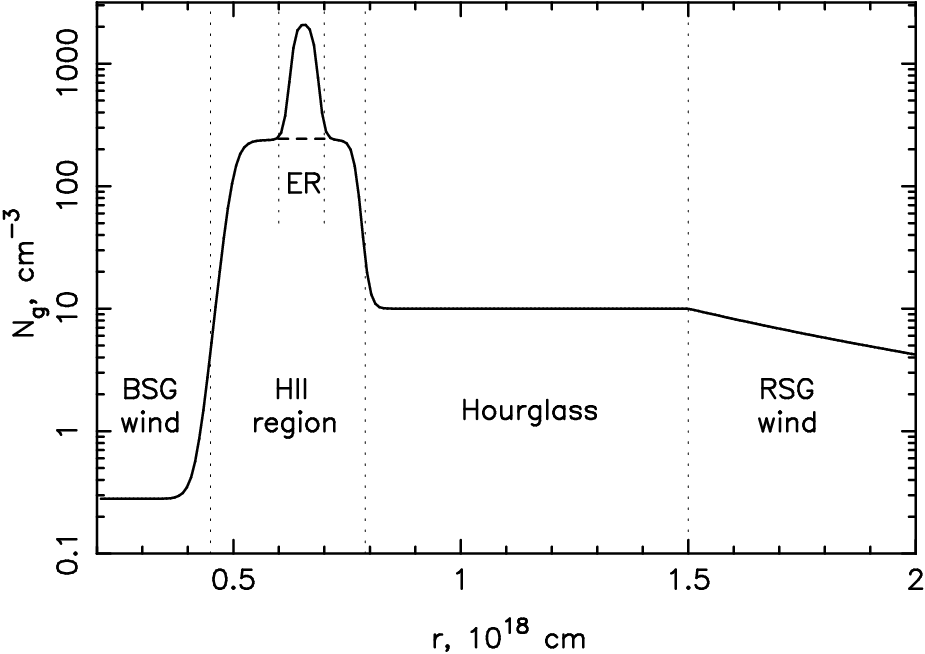} 
\figcaption{Radial profiles of CSM gas density $N_\mathrm{g}$. The partly 
overlapping thick dashed and solid lines, respectively, correspond to region 1 
($4.5^{\circ}<|\theta|<20^{\circ}$) and region 2 ($|\theta|<4.5^{\circ}$) in 
elevation angle $\theta$. The radial extent of the various morphological 
structures, summarized in section 2, is indicated by the vertical dotted lines.
\label{f1}}
\end{figure}

The propagation of the forward SN shock through the CSM is modeled in the spirit 
of a spherically symmetrical approach. It approximates the shock and its effects 
as the weighted sum of two independent spherically symmetric shocks, propagating 
into, respectively, two different radial gas number density profiles of the CSM, 
shown in Figure~\ref{f1}: the first profile (region 1) belongs to an azimuthally 
symmetric region of elevation angles $\theta$ near the equatorial plane 
$4.5^{\circ}<|\theta|<20^{\circ}$, and the second, analogous profile (region 2), 
corresponds to the innermost $|\theta|<4.5^{\circ}$. Then, each quantity Q, 
characterizing the number of accelerated CRs and the amount of emission 
produced by these CRs, is determined by the relation
\begin{equation}
Q=Q_1(f_1-f_2)+Q_2f_2,
\end{equation}
where $Q_{1,2}$ are the spherically symmetric values corresponding to
 the profiles that characterize the two regions 1 and 2, respectively, and
$f_1=0.34$ and $f_2=0.08$ denote the filling factors in the solid angle
of these regions.
%
%

CR production by the SN shock at high latitudes $|\theta|>20^{\circ}$ is
 neglected here because the gas density in this region is considerably lower
(see Potter14). We also neglect CR production at the reverse shock for the 
reasons given in BKV11.

The question is, of course, to what extent such an approximate treatment remains
consistent as a function of time with the real SNR shock sweeping across the
real CSM to reach larger and larger radial distances. A priori the decrease of
the gas density with elevation angle, symmetric to the equatorial plane,
suggests this to be a roughly stable process. However, effects like the
engulfment of the ER clearly imply some non-radial shock propagation
aspects. Such effects are also apparent in the three-dimensional simulations of
Potter14. In addition, the ring is clumpy, even though \citet{fransson15}
found indications that these hot spots are now gradually dissolving. This
type of effect should primarily influence the detailed time dependence of the
particle acceleration and hadronic \gr emission rather than the
global acceleration properties of the system\footnote{For discussions of
such effects, however, see \citet{bkv13} and \citet{gabici14}.}.
Therefore, in a gross sense, the used mosaic of spherical shocks appeears
to be an adequate overall approximation.

\subsection{Field Amplification, Injection Rate, and Electron to Proton Ratio}

As in BKV11, the magnetic field strength, $B_0$, given by the expression
\begin{equation}
B_0=\sqrt{2\pi \times 10^{-2}\rho_0 V_\mathrm{s}^2}
\end{equation}
is used. Here, $V_\mathrm{s}$ denotes the SN shock speed and $B_0$ is the field far
upstream, presumably amplified by the CRs of the highest energy. In the
same sense, $\rho_0$ is the mass density far upstream. The high downstream
magnetic field $B_{\mathrm d} = B_0 \times \sigma \approx 10$~mG, where $\sigma$
denotes the total shock compression ratio, is required to reproduce the
observed radio and X-ray spectra \citep{bk06}. The calculation of the
  shock radius $R_\mathrm{s}$ and speed $V_\mathrm{s}$
  again follows the scheme of \citep{bk06}, but see also BKV11, as does the
  evaluation of the proton injection rate $\eta N_\mathrm{g}$ and of the
  electron-to-proton ratio $K_\mathrm{ep}$.

\begin{figure}
\plotone{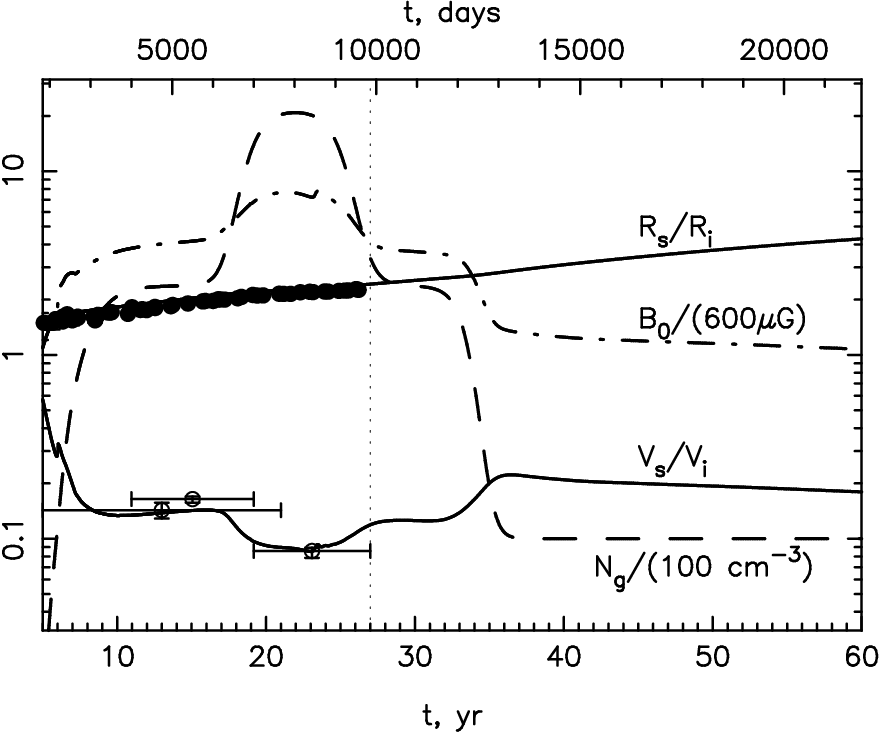} 
\figcaption{Shock radius $R_\mathrm{s}$ and shock speed $V_\mathrm{s}$ 
    (solid lines), gas number density $N_\mathrm{g}$ (dash-dotted line)
  and upstream magnetic field $B_0$ (dashed line) at the current shock
  position as a function of time since SN explosion, for region 2. The 
  dotted vertical line marks the current epoch. The observed radius
  $R_\mathrm{s}$ and speed $V_\mathrm{s}$ of the SN shock, as determined by
  radio observations \citep{ng13} are shown as well. The
  scaling values are $R_\mathrm{i}=R_\mathrm{T}=3.1\times 10^{17}$~cm and
  $V_\mathrm{i}=28000$~km~s$^{-1}$.
\label{f2}}
\end{figure}

\section{Results and Discussion}

\begin{figure}[t]
\plotone{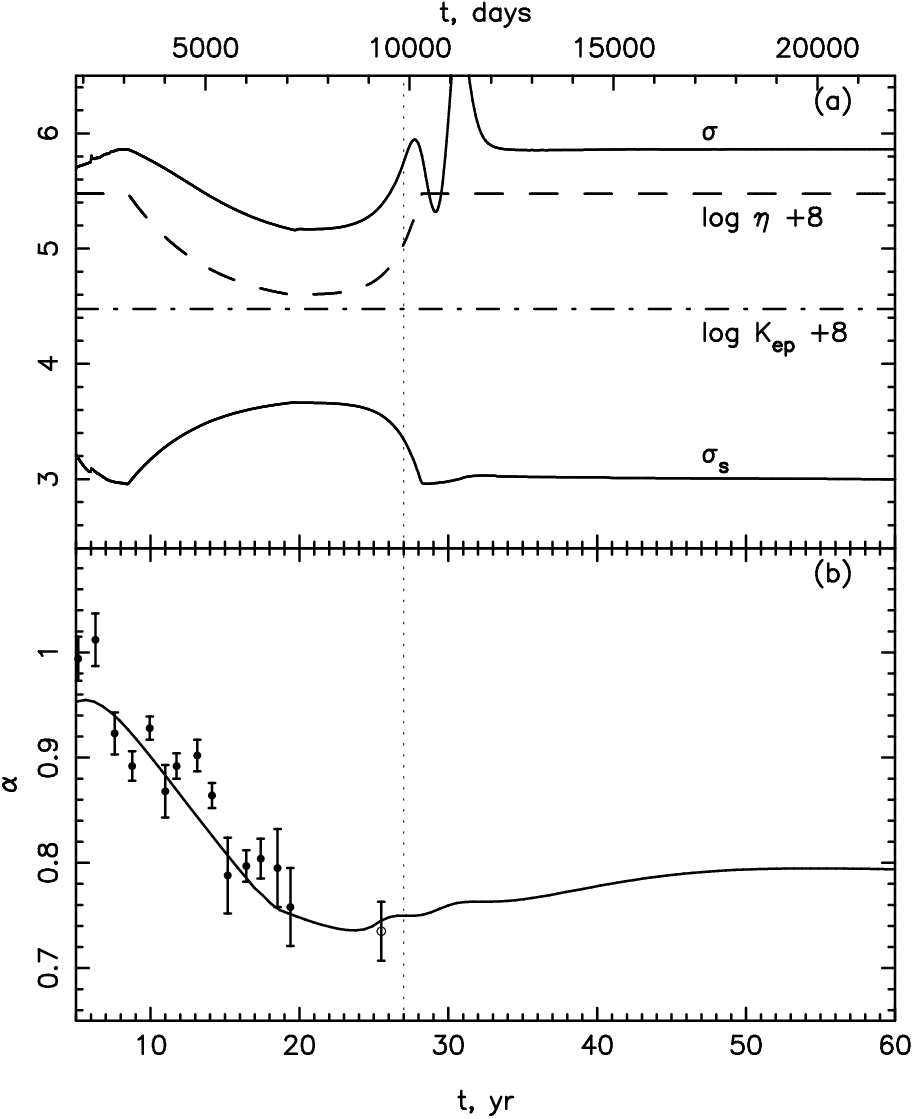} \figcaption{(a) Shock compression ratio $\sigma$ and subshock 
compression ratio $\sigma_\mathrm{s}$ (solid lines), proton injection 
fraction $\eta$ (dashed line), and electron-to-proton ratio 
$K_\mathrm{ep}$ (dash-dotted line) as functions of time. (b) 
Self-consistent spectral index $\alpha$ of the integrated nonthermal radio 
emission as a function of time together with observational data from ATCA 
\citep{zanardo10} and the latest combination of ATCA and ALMA 
\citep{zanardo14}.
\label{f3}}
\end{figure}

Figure~\ref{f2} shows $R_\mathrm{s}$ and $V_\mathrm{s}$, as the shock
propagates in the CSM corresponding to region 2, together with the latest radio
data \citep{ng13}. In the case of region 1, the shock speed time profile
$V_\mathrm{s}(t)$ does not contain the local minimum around $t = 8000$ days; this
is the main difference from the results presented in Figure~\ref{f2}. Iteratively
fitting the theoretical quantities $\eta(t)$ and $K_\mathrm{ep}(t)$ to the
spatially integrated radio synchrotron spectra up to the year 2013
\citep{ng13} leads to a constant value for $K_\mathrm{ep}(t)= 3\times
10^{-3}$, whereas the value $\eta(t)\approx 3\times 10^{-3}$ at $t\approx
26$ year (Figure~\ref{f3}a) is due to the assumption that, leaving the H~II
region, the nuclear injection fraction should, after $t\approx 30$ year, go back
to its value before the age of 10 years. During the ages between about 10
and 30 years the compressed and largely azimuthal magnetic field of the H~II
region should have depressed nuclear injection, consistent with the softening
of the spatially integrated radio spectrum (with index $\alpha(t)$) as shown in
Figure~\ref{f3}b. This figure also shows that outside the H~II region, the
shock is significantly modified relative to a pure gas shock for which the
total shock compression ratio $\sigma$ and the subshock compression ratio
$\sigma_\mathrm{s}$ would both have a value of 4.

%

The adopted value of $\eta(t)$ determines the amount of shock modification, in
particular, the decrease of the subshock compression ratio $\sigma_\mathrm{s}(t)$
relative its value of 4 for an unmodified shock. Since CRs with relatively
small energies are produced near the subshock, the value $\sigma_\mathrm{s}(t)$ directly
determines the shape of the electron energy spectrum at energies below $\approx
1$~GeV --- which produce synchrotron emission in the radio range ---  and vice
versa. Therefore, as in all similar cases, the proton injection rate
$\eta(t)$ is inferred from a fit of the resulting theoretical spectral index
$\alpha(t)$ of the integrated radio synchrotron emission to the observed
value. The quality of this fit can be ascertained from Figure~\ref{f3}b, where 
the best calculated time profile $\alpha(t)$ is presented together with the 
observational data.

From the epoch $t>6$ year onwards the value of $\alpha(t)$ decreases until the epoch
$t\approx 23$ year \citep{zanardo10} and is assumed to start to increase again at
$t> 23$ year. This appears to be consistent with recent measurements by
  \citet{indeb14} and, especially, \citet{zanardo14}. However, the increase is
not as fast as one would expect from the behavior of the local
$\sigma_\mathrm{s}(t)$. This occurs because at subsequent epochs, the gas number density
and, consequently, also the amount of freshly injected electrons and the
magnetic field strength, are much smaller than when the shock was in the H~II
region. Therefore, the synchrotron emission of those latter electrons, convected
downstream in that higher magnetic field, remains dominant. For similar
reasons, the variation in the shock compression ratio $\sigma$ at an age
10,000--12,000 days, caused by the rapid decrease of the
upstream gas number density $N_\mathrm{g}$, does not significantly affect the
integrated radio synchrotron emission flux, and the spectral index $\alpha$
remains essentially independent of $\sigma$.

\begin{figure}
\plotone{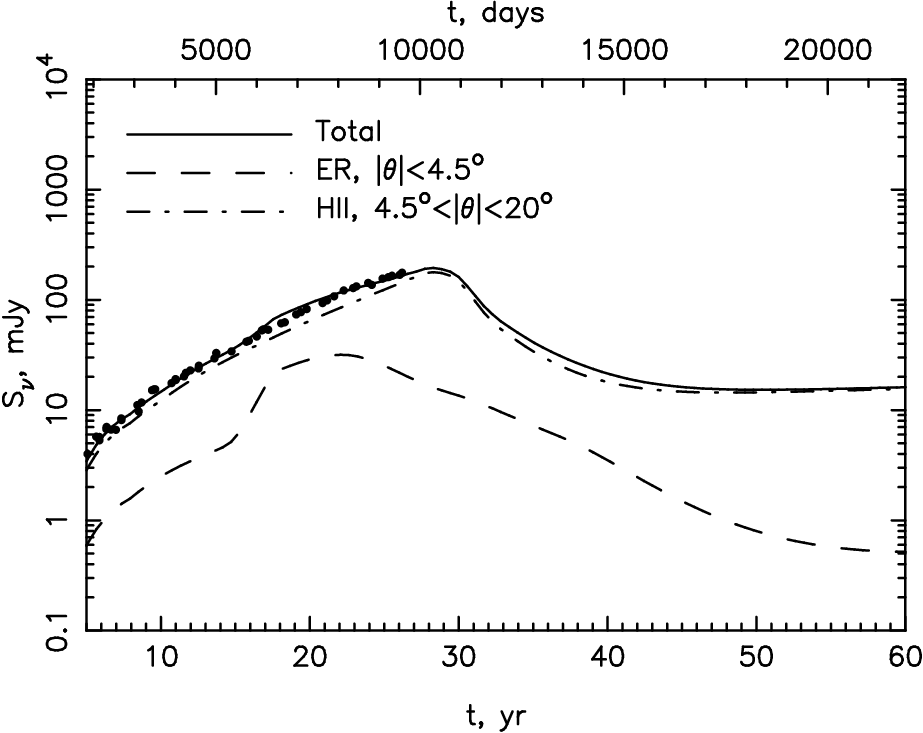} \figcaption{Calculated flux of radio emission at frequency $\nu 
= 9$~GHz as a function of time, together with the ATCA data \citep{ng13}. Dashed 
and solid lines represent the contribution of region 1 and of the sum from 
regions 1 and 2, respectively.
\label{f4}}
\end{figure}

Using these educated guesses for $\eta$ and $K_\mathrm{ep}$ the calculated flux
of radio emission $S_{\nu}$ at frequency $\nu = 9$~GHz is presented in
Figure~\ref{f4} together with the observational data obtained with the ATCA
instrument \citep{ng13}. For those times where the radio flux has also been
measured, the calculation is perfectly consistent with the
observations. According to the calculation, the rapid growth of radio emission
in the epochs $t<30$ year is due to the increase of the number of accelerated CR
electrons, which is proportional to the swept-up mass within the H~II region.
For the adopted value of the outer boundary of the H~II region, $R_\mathrm{HG}=
8\times10^{17}$~cm, the SNR shock reaches this boundary after $t\approx 30$ year,
and the peak of radio emission is reached. Even at later epochs, $t=30-40$ years,
the radio synchrotron emission will still be dominated by the contribution of
the swept-up matter of the H~II region. The flux of radio emission is expected
to decrease gradually by a factor of about 10 during the 10 years after 2017. Due
to its considerably lower gas density, the contribution of the swept-up
hourglass matter will become essential only for $t>43$ year, when the radio
emission starts to grow slowly again.

\begin{figure}[t]
\plotone{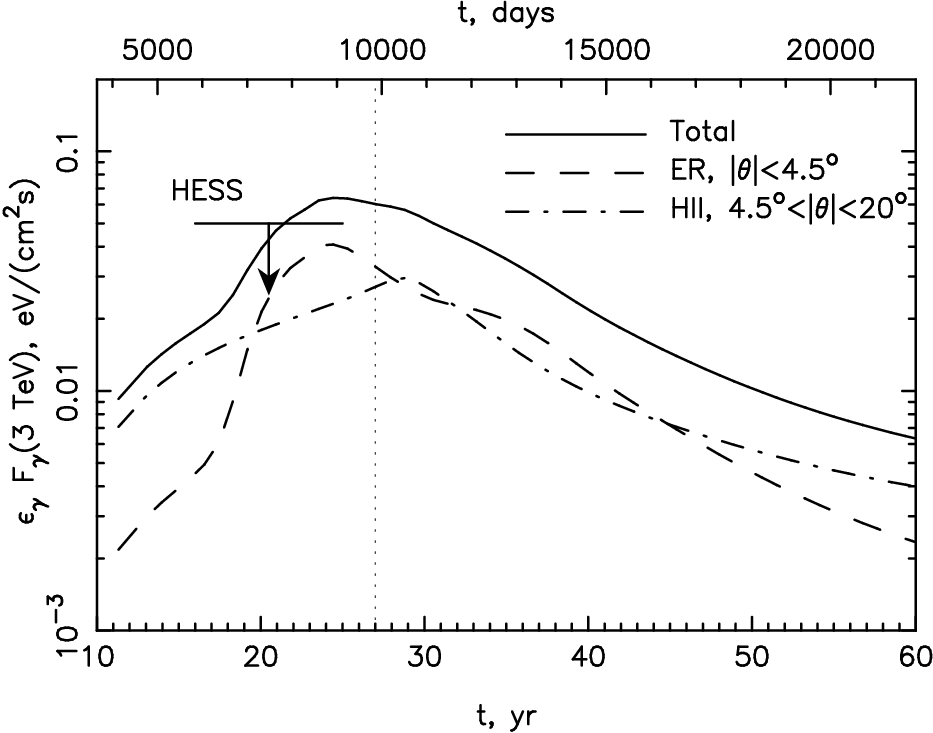} \figcaption{Integral \gr energy flux density above 3 TeV from 
SN~1987A as a function of time. The H.E.S.S. \citep{hess15} upper limit, 
corresponding to the observational period 2005--2012 year, is shown as well.
\label{f5}}
\end{figure}

The calculated \gr energy flux density above 3 TeV as a function of time, shown in
Figure~\ref{f5}, is dominated by the $\pi^0$-decay component at all energies.
Since a significant part of the shock surface is expected to be tangential, and
therefore to not efficiently inject/accelerate nuclear CRs, the overall number
of accelerating CRs is normalized by a factor of $f_\mathrm{re}=0.2$, as previoously 
argued by \citet{vbk03} and BKV11.

As is clear from Figure~\ref{f5}, the region 1, which contains the ER, the
densest structure, contributes dominantly only during the shock propagation
through the ER, which is during days 7000--10,000.

According to Figure~\ref{f5}, the maximal energy flux density of TeV emission
$\epsilon_{\gamma}F_{\gamma}\approx 7\times 10^{-2}$~eV cm$^{-2}$ s$^{-1}$ was
achieved at day 9000, and after that epoch it decreased continuously due to the
decrease of the CSM gas density. Since the radial gas density profile is much
thinner compared with the one used earlier BKV11, the peak value of the
expected flux of TeV emission is now lower and the TeV emission expected for
the future is considerably lower.

The most recent upper limit for the TeV emission obtained by the H.E.S.S.
telescopes during the period 2005-2012 \citep{hess15} (see Figure~\ref{f5}) is
roughly consistent with this prediction.

According to the calculation (see Figure~\ref{f5}), the most promising time for 
the detection of SN~1987A in TeV \grs is the 10 year period from 2008 to 2018. 
At later epochs, SN~1987A should be detectable in the VHE range only by an 
instrument with a higher sensitivity than that of H.E.S.S.

\acknowledgements This work has been supported in part by the Russian
Foundation for Basic Research (grants 13-02-00943 and 13-02-12036) and by the
Council of the President of the Russian Federation for Support of Young
Scientists and Leading Scientific Schools (project No. NSh-3269.2014.2).

\section{Appendix}
\begin{figure}[h]
\plotone{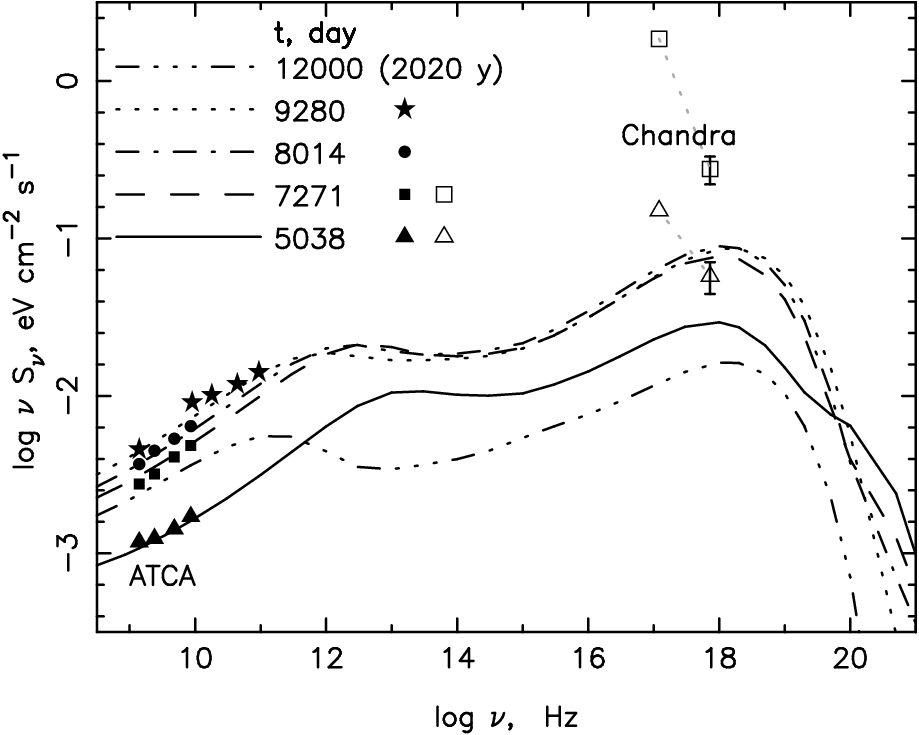} \figcaption{Spatially integrated synchrotron spectral energy 
flux density of SN 1987A, calculated for five evolutionary epochs. The ATCA 
radio data \citep{zanardo10,zanardo14} for four epochs are shown as well, 
together with the {\it Chandra} X-ray flux \citep{park2007} data for 
two epochs (both in the soft energy range at $\nu = 10^{17}$ Hz and, connected 
by dotted lines, in the hard energy range at $\nu = 6 \times 10^{17}$ Hz).
\label{a1}}
\end{figure}
\begin{figure}[h]
\plotone{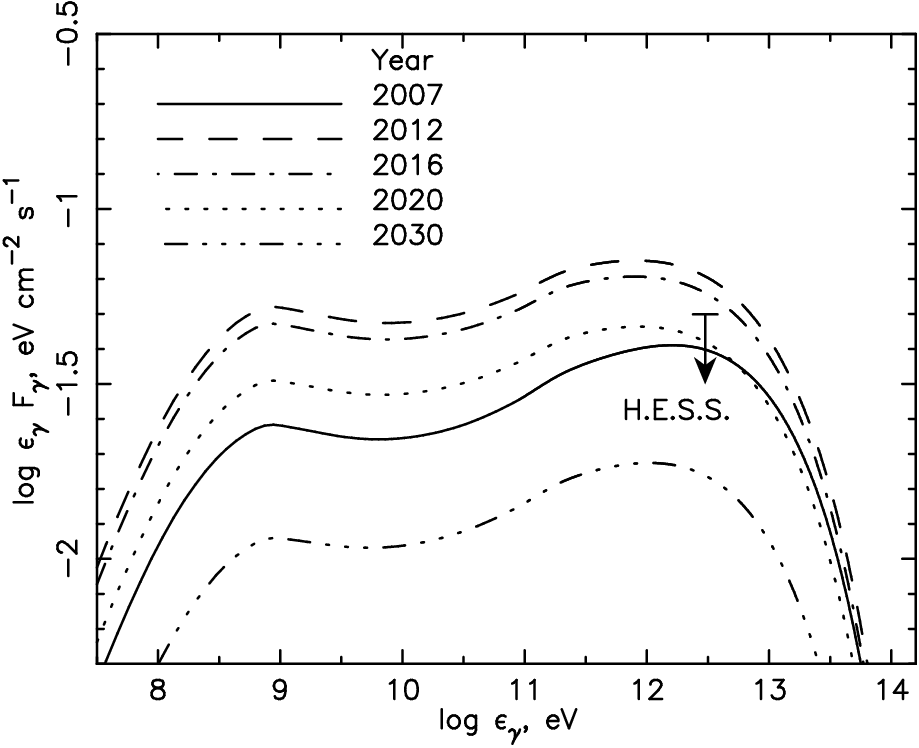} \figcaption{Spatially integrated $\gamma$-ray spectral energy 
flux density, calculated for five epochs. The recent H.E.S.S. upper limit 
\citep{hess15} is shown as well.
\label{a2}}
\end{figure}

\end{document}